\documentclass[aps,prb,twocolumn,superscriptaddress]{revtex4-1}

\usepackage{hyperref} 
\usepackage{graphicx}
\usepackage{amsmath}
\usepackage{amssymb}

\DeclareGraphicsExtensions{.png,.jpg,.eps}
\usepackage{xcolor}

\begin{document}

\author{D. Soriano}
\affiliation{Radboud University, Institute for Molecules and Materials, NL-6525 AJ Nijmegen, the Netherlands}

\author{J. L. Lado}
\affiliation{Department of Applied Physics, Aalto University, 00076 Aalto, Espoo, Finland}

\title{
	Exchange-bias controlled
	correlations
	in magnetically encapsulated
twisted van der Waals dichalcogenides
}

\begin{abstract}
Twisted van der Waals materials have become a paradigmatic platform to realize exotic correlated states of matter. Here, we show that a twisted dichalcogenide bilayer (WSe$_2$) encapsulated between a magnetic van der Waals material (CrBr$_3$) features flat bands with tunable valley and spin flavors.  We demonstrate that, when electron-electron interactions are included, spin-ferromagnetic and valley-ferromagnetic states emerge in the flat bands, stemming from the interplay between correlations, intrinsic spin-orbit coupling and exchange proximity effects. We show that the specific symmetry broken state is controlled by the relative alignment of the magnetization of the encapsulation, demonstrating the emergence of correlated states controlled by exchange bias. Our results put forward a new van der Waals heterostructure where symmetry broken states emerge from a genuine interplay between twist engineering, spin-orbit coupling and exchange proximity, providing a powerful starting point to explore exotic collective states of matter.  
\end{abstract}

\date{\today}

\maketitle

\section{Introduction}

Two-dimensional materials have risen as a paradigmatic
platform
to engineer emergent states of matter. This flexibility stems from
the capability of stacking different
two-dimensional materials on top of each
other, allowing to combine electronic orders.
Among the different two-dimensional materials, 
transition metal dichalcogenides\cite{Radisavljevic2011,Splendiani2010,Manzeli2017,PhysRevLett.105.136805,Dickinson1923,Wilson1969}
attracted much attention due to their strong spin orbit-coupling effects.
This has allowed exploiting different early proposals in spintronics
such as spin-charge conversion\cite{PhysRevLett.119.196801} 
valley control,\cite{Zeng2012,Safeer2019,Shao2016,Mendes2018} and optically generated spin currents,\cite{Luo2017,Avsar2017}
demonstrating the possibilities for magnetic control in two-dimensional
materials. In this line, the recent discovery, and isolation
of two-dimensional magnetic materials\cite{Huang2017,Gibertini2019,Samarth2017,Gong2017,Lin2016,Du2015,Kuo2016} 
has allowed pushing this idea
even further, providing a purely van der Waals platform for spintronic
physics.

Moire structures are known to emerge in generic twisted two dimensional
materials, including graphene,\cite{PhysRevB.86.155449,PhysRevLett.99.256802}
transition metal dichalcogenides (TMDC),\cite{PhysRevLett.122.086402,PhysRevLett.121.266401,PhysRevResearch.2.013335,PhysRevB.91.165403,Jones2013,Baugher2014}
boron nitride,\cite{PhysRevLett.124.086401} MoO$_3$\cite{2020arXiv200414588C}
and ferromagnets.\cite{Hejazi2020}
In particular, twisted transition metal dichalcogenide systems provide
unique opportunities for spintronics, due to the strong spin-orbit
coupling.\cite{PhysRevX.4.011034,Mak2012,Zeng2012,Cao2012,PhysRevB.86.081301,Xu2014,Mak2016}
Correlated states in twisted dichalcogenides
materials have been also demonstrated,
stemming from the emergence of quasi-flat
bands.\cite{2019arXiv191012147W,PhysRevLett.121.026402,PhysRevLett.121.266401,2019arXiv190810399N,PhysRevLett.122.086402,2020arXiv200103812W,Jin2019TM,Regan2020TM,Shimazaki2020,Shimazaki2020,Tang2020TM,2019arXiv191014061Z,2020arXiv200313690S,2020arXiv200513868Z,2019arXiv190104679F,2019arXiv191013068Z,2020arXiv200513879Z}
Importantly, the
strong spin-orbit effects, together with the emergent
valley and layer degrees of freedom, bring forward
a whole new set of possibilities\cite{Manzeli2017,Xu2014,Mak2016,Hill2016} for engineering
strongly correlated states of matter
in twisted dichalcogenides. In particular,
the spin-valley locking\cite{PhysRevB.84.153402,PhysRevLett.108.196802}
provides a way of controlling the internal
quantum numbers of a symmetry broken state.

\begin{figure}[t!]
\centering
    \includegraphics[width=\columnwidth]{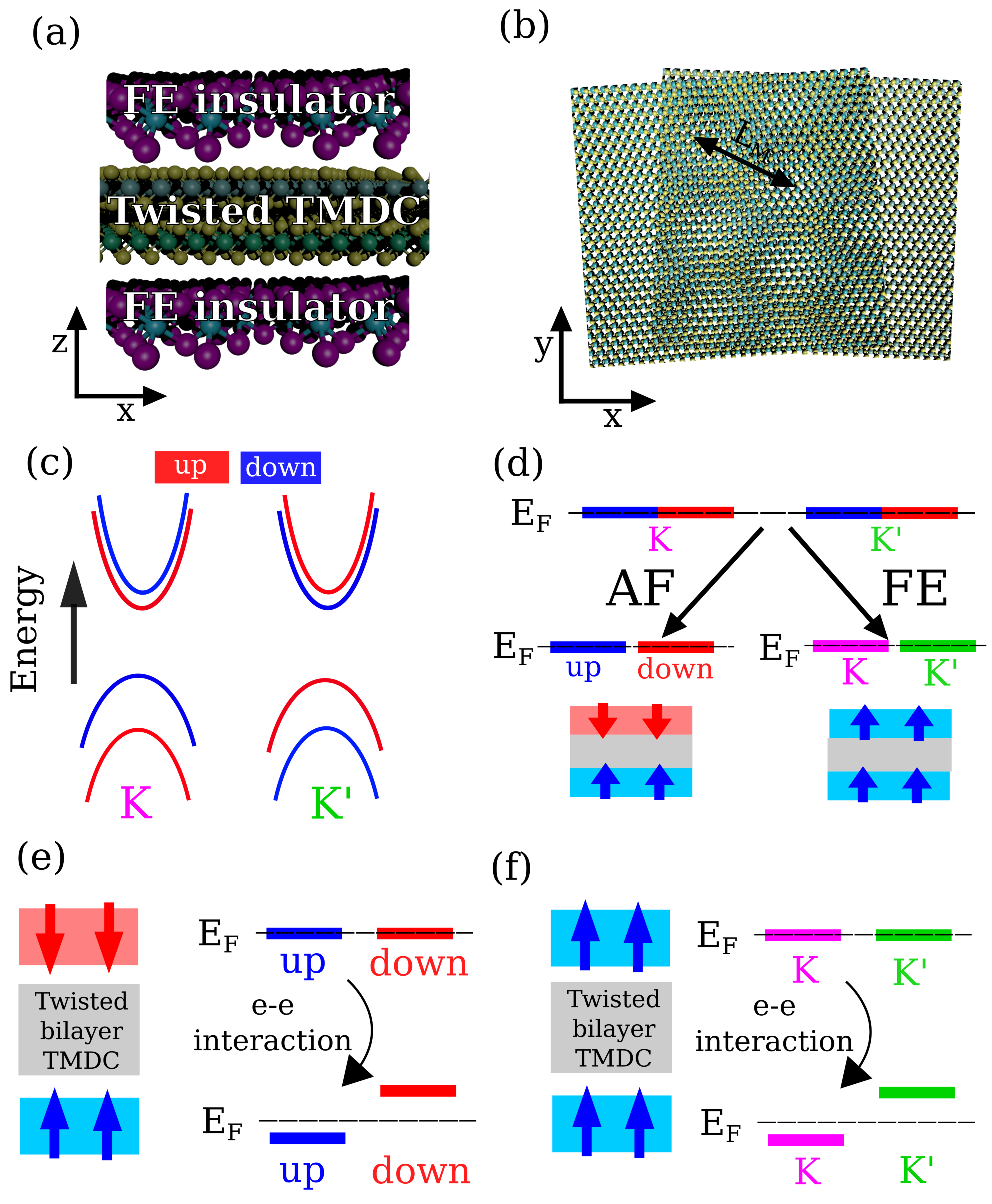}

\caption{
	Sketch of the van der Waals exchanged dichalcogenide multilayer (a) 
	and sketch of the emergence of moire superlattice (b). Panel
	(c) shows a sketch of the typical band structure of TMDC when
	spin-orbit coupling effects are considered.
	Panel (d) shows the possible lifting of degeneracies that
	the magnetic encapsulation can create.
	Panel (e,f) shows the interaction-induced symmetry breaking
	of the low energy states. For antiferromagnetic alignment (e)
	spin symmetry broken state appears, whereas for ferromagnetic alignment (f)
	a valley symmetry broken state emerges.
}
\label{fig:fig1}
\end{figure}

Here we show that a correlated state controlled by exchange bias
can be engineered
in a van der Waals heterostructure
made of magnetic two-dimensional materials and a
twisted dichalcogenide bilayer. In particular, we demonstrate
that a magnetic exchange bias can control the symmetry broken
state of twisted dichalcogenide system, creating
a transition from valley ferromagnetism to spin ferromagnetism.
Our results put forward exchange bias as a powerful knob to
control symmetry broken states in 
magnetically encapsulated twisted
dichalcogenide materials.
The manuscript is organized as follows. 
In Section \ref{sec:bands} 
we
show the emergence of nearly flat bands
controlled by spin-orbit coupling and exchange coupling
in the magnetically encapsulated multilayer.
In Section \ref{sec:int} we show the selective switching
between interaction-induced
valley ferromagnetism and spin ferromagnetism
by controlling the magnetization of the
encapsulation.
In Section \ref{sec:ex} we estimate the strength of the effective
exchange fields by means of first-principles calculations
for a specific CrBr$_3$/WSe$_2$ based structure.
Finally, in Section \ref{sec:int} we
summarize our conclusions.

\begin{figure}[t!]
\centering
    \includegraphics[width=\columnwidth]{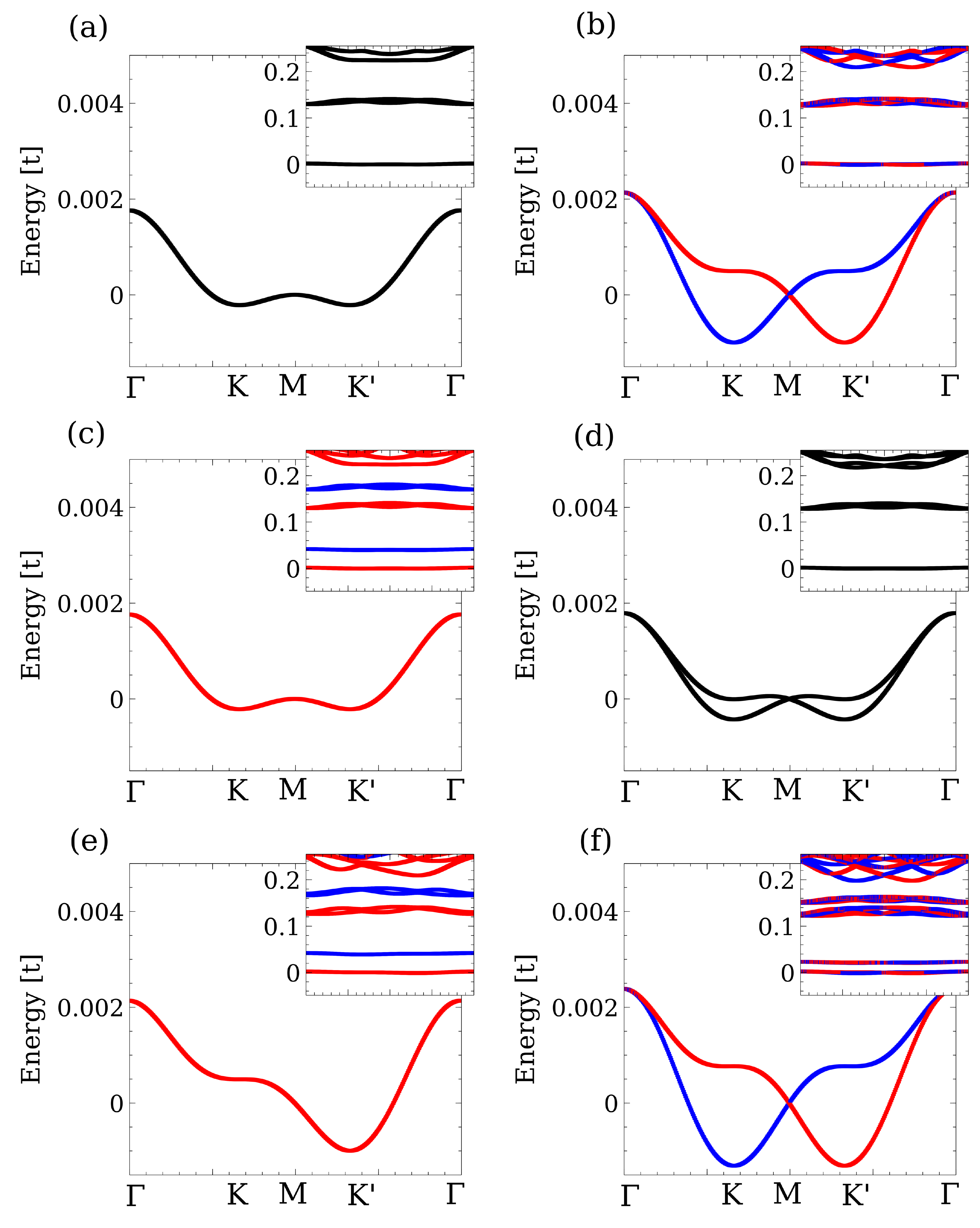}

\caption{
	Band structure of a twisted dichalcogenide
	with a twist angle of $1.8^\circ$ in the absence of
	exchange effects (a,b), with zero SOC (a) and
	with finite SOC (b).
	Panels (c,d) shows the bandstructure in the absence
	of SOC, but with a finite exchange bias,
	having a ferromagnetic alignment
	between the layers in (c) and antiferromagnetic in (d)
	Panels (e,f) show the band structures with finite
	SOC and exchange fields, 
	having a ferromagnetic alignment
        between the layers in (e) and antiferromagnetic in (f).
	In the cases (e,f), the low energy states are
	formed by two bands, yet with dramatically
	different orbital content as sketched in
	Fig. \ref{fig:fig1}d.
}
\label{fig:fig2}
\end{figure}

\section{Controlling flat bands with exchange bias}
\label{sec:bands}
We first address the emergence of flat bands in the magnetic van der Waals
heterostructure, and in particular how spin-orbit and exchange effects
the low energy flat bands.
In the following, we consider an effective model
for a transition metal dichalcogenide in a honeycomb lattice, 
that captures
in an effective
fashion the low energy valence and conduction bands.
For the sake of concreteness, we will focus on the physics emerging in the
valence band of the model.
We will consider a twisted van der Waals dichalcogenide, encapsulated
between a ferromagnetic insulator, where we have integrated out
the degrees of freedom of the ferromagnetic insulator.
The total Hamiltonian is of the form

\begin{equation}
	H = H_{0} + H_{SOC} + H_{J}
\end{equation}

with $H_0$ the non-relativistic Hamiltonian, $H_{SOC}$ the spin-orbit
coupling correction and $H_J=H_{AF}/H_{FE}$ the exchange proximity effect

	\begin{multline}
	H_0 = \sum_{\langle ij \rangle} t c^\dagger_{i,s} c_{j,s}
	+ m\sum_{i,s} \vartheta^z_{ii} c^\dagger_{i,s} c_{i,s}  \\
	+ \sum_{i,j,s} t^\perp (\mathbf{r}_i,\mathbf{r}_j) c^\dagger_{i,s} c_{j,s}
\end{multline}
\begin{equation}
	H_{SOC} = i\lambda_{SOC} \sum_{\langle\langle ij \rangle\rangle,s,s'} 
	\nu_{ij}
	\sigma^z_{s,s'} c^\dagger_{i,s} c_{j,s'}
\end{equation}

\begin{equation}
	H_{AF} = J_{AF} \sum_{i,s,s'}
	\tau_i
	\sigma^z_{s,s'} c^\dagger_{i,s} c_{i,s'}
\end{equation}

\begin{equation}
	H_{FE} = J_{FE} \sum_{i,s,s'}
	\sigma^z_{s,s'} c^\dagger_{i,s} c_{i,s'}
\end{equation}

where $t$ is a first neighbor hopping, $t^\perp (\mathbf{r}_i,\mathbf{r}_j)$ parametrize
the interlayer coupling,\cite{PhysRevB.82.121407,PhysRevB.92.075402,PhysRevLett.119.107201} 
$\vartheta^z_{ii}$ is the sublattice Pauli matrix,
$m$ denotes the onsite energy imbalance between the two sites
of the honeycomb lattice,
$\sigma^z_{s,s'}$ is the spin Pauli matrix,
$\nu_{ij}=\pm 1$ 
for clock-wise/anti-clockwise
hopping,\cite{PhysRevLett.95.226801}, $\langle \rangle$ denotes
sum over first neighbors in a layer,
$\langle \langle \rangle \rangle$ denotes sum over second
neighbors in a layer,
and $\tau_i=\pm 1$ for upper/lower layer.
The previous model applied for a single layer faithfully
captures the electronic structure
at the $K$ and $K'$ points,\cite{PhysRevLett.108.196802,PhysRevB.88.085440,PhysRevB.87.155304} and
therefore can be used as a starting point for flat bands
stemming from the dichalcogenide valleys.
We note that more sophisticated models for a dichalcogenide could
be considered, yet for the sake of simplicity we here focus on
the minimal one that captures the spin-valley physics at the
valleys.

\begin{figure}[t!]
\centering
    \includegraphics[width=\columnwidth]{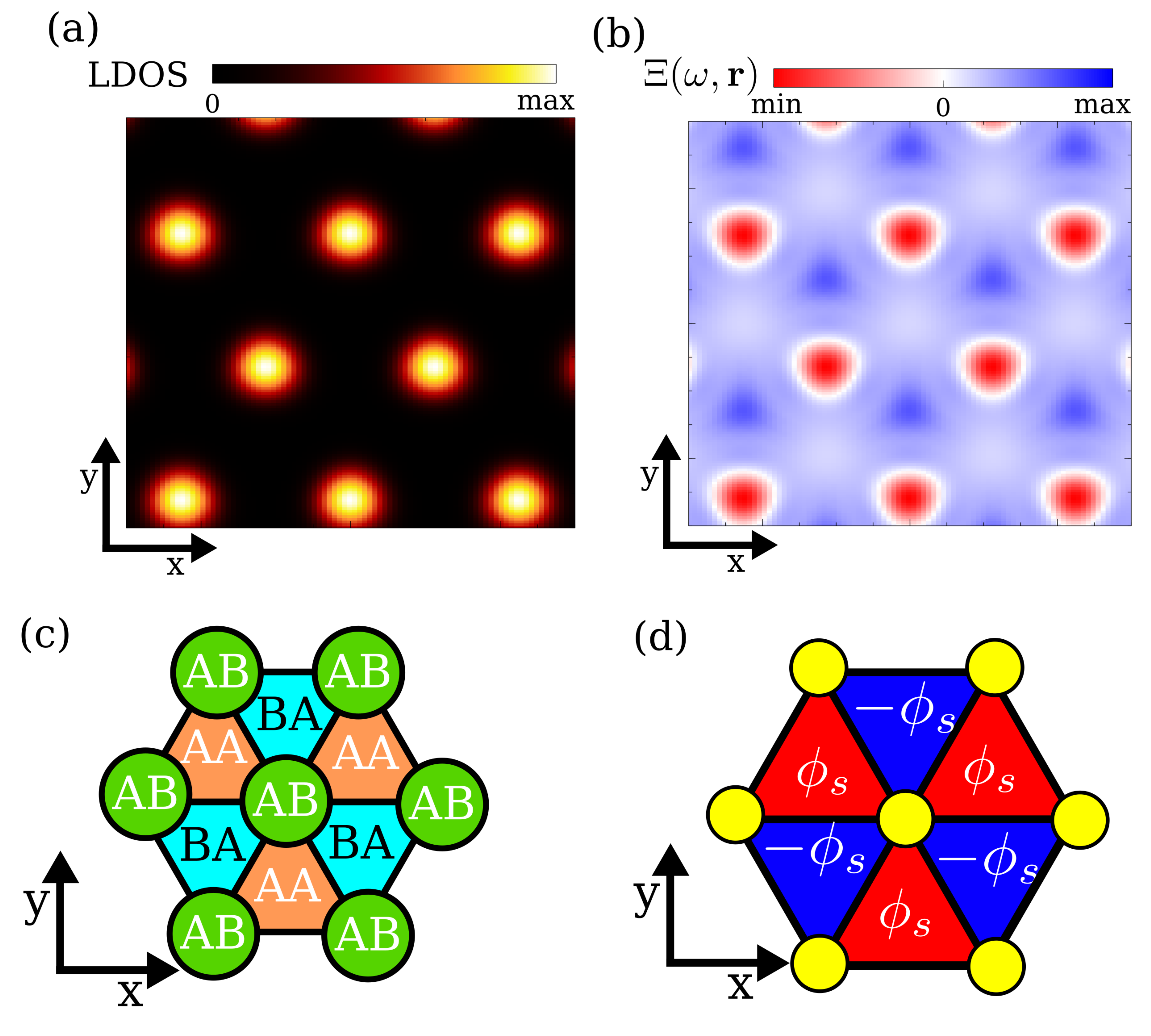}

\caption{Local density of states of the lowest flat band, showing
the emergence of localized states in a triangular lattice.
Focusing now in the heterostructure with antiferromagnetic alignment
	of Fig. \ref{fig:fig2}f,
	panel (b) shows the real space spin flux
	at $\omega=\epsilon_F$, showing the emergence
	of non-trivial spin topology.
	Panel (c) shows the sketch of the different parts of the superlattice,
	and panel (d) a low energy model for the nearly flat bands.
}
\label{fig:fig3}
\end{figure}

With the previous effective model, we now compute the electronic structure
of the magnetically encapsulated dichalcogenide multilayer. In particular,
it is interesting to look in detail at the degeneracy of the low
energy manifold in different regimes.
Let us start with the case $J_{AF}=J_{FE}=\lambda_{SOC}=0$, the case in which
both exchange and spin-orbit coupling effects are neglected.
As shown in 
Fig.
\ref{fig:fig2}a the band structure for a structure
with a twist angle $1.8^\circ$ shows
nearly flat bands as expected from twisted dichalcogenide systems.
It is important to note that such energy bands are four-fold degenerate,
two times from valley, and two times due to spin. Such phenomenology
is analogous to twisted graphene multilayers or twisted boron nitride.
Due to the existence of such degeneracy, it is expected that the addition
of additional terms breaking that symmetry will create splittings 
between the different sectors.
When we consider the effect of intrinsic spin-orbit
coupling ($\lambda_{SOC}=0.02t$)
as shown in 
Fig. \ref{fig:fig2}b a momentum-dependent
spin splitting appears, yet the valley degeneracy remains along the
shown k-path. 
As a result, the low energy manifold remains four-fold degenerate.
It is interesting to note that the inclusion of spin-orbit
coupling slightly increases the bandwidth of the low energy bands, yet
still keeping an overall small bandwidth. Besides the splitting created
in the low energy bands, additional splittings are created at higher energies.

Let us now focus in the cases hosting only exchange fields, and
in the absence of spin-orbit coupling. The simplest case to consider
is to the ferromagnetic configuration ($J_{FE}=0.02t$), in which
the spin degeneracy is lifted Fig. \ref{fig:fig2}c. In this situation,
a uniform shift between up and down channels take place,
and the low energy bands become just two-fold degenerate due to
valley degeneracy. A more interesting scenario happens for
antiferromagnetic exchange bias ($J_{AF}=0.05t$) Fig. \ref{fig:fig2}d,
in which a small splitting appears in the bands, yet without fully
lifting the degeneracy of the low energy manifold.
It is especially interesting to note that the band structure
with spin-orbit coupling (Fig. \ref{fig:fig2}b) 
and antiferromagnetic exchange 
bias
(Fig. \ref{fig:fig2}d) both show a four-fold
degeneracy of the low energy manifold, in contrast
with the trivial lifting of degeneracy created by the
ferromagnetic exchange field (Fig. \ref{fig:fig2}c).

We now move on to the case when both exchange fields and spin-orbit
coupling are non-zero.
We start with the case of ferromagnetic exchange
($J_{FE}=0.02t$) and non-zero SOC ($\lambda_{SOC}=0.02t$), 
shown in Fig. \ref{fig:fig2}e. The low energy manifold
is still two-fold degenerate stemming from the valley
degree of freedom (both bands are
superimposed in Fig. \ref{fig:fig2}e), 
yet with a clear asymmetry
between $-k$ and $+k$ due to the combination of SOC effect
and exchange field. A more interesting scenario corresponds to the
antiferromagnetic exchanged case
($J_{AF}=0.02t$) with non-zero SOC ($\lambda_{SOC}=0.02t$)
shown in Fig. \ref{fig:fig2}f. It is clearly observed that now the
low energy manifold is two-fold degenerate, in contrast with
Figs. \ref{fig:fig2}bd (see the splitting
showed in the insets). The low energy degeneracy comes from a combination
of time-reversal and mirror symmetries, and could be lifted with an
electric field between the layers. It is worth to remark that in both cases,
ferromagnetic and antiferromagnetic bias, the low energy manifold
is two-fold degenerate, yet with dramatically different quantum numbers.
This feature will be important when considering the effect of electronic 
interactions.

Before moving on to interaction effects,
it is interesting to understand the nature
of the low energy states (Fig. \ref{fig:fig3}abcd).
The states associated to the two low energy flat bands form an emergent
triangular lattice, as shown in the local density of states of
(Fig. \ref{fig:fig3}a). 
This is of course already suggested by the band structures
of Fig. \ref{fig:fig2}, that features the typical dispersion of a 
triangular lattice.
Interestingly, the location of this emergent triangular
superlattice corresponds to regions of the twisted system 
showing AB stacking (Fig. \ref{fig:fig3}c),
namely a region where one of the atoms fall
in the hollow site, 
in strike contrast with the conventional AA localization in twisted
graphene bilayers.\cite{PhysRevB.82.121407,PhysRevB.88.121408,Bistritzer2011} 
This is easily rationalized by taking into account that,
whereas for graphene AB and BA regions are equivalent, in a twisted
dichalcogenide bilayer AB corresponds to Mo-Mo stacking, whereas
BA corresponds to S-S stacking, and therefore are inequivalent. As a result,
a triangular arrangement of the states in the AB regions becomes possible.

The triangular arrangement between the states, together with the low energy
dispersion, suggests that this system can be described with a low
energy effective model. To gain insight into this, it is interesting
to look at the valley spin 
flux\cite{PhysRevB.84.205137,2020arXiv200305163M,PhysRevLett.123.096802,2020arXiv200502169L} in the moire unit cell defined as

\newcommand{\gvz}{\mathcal{G}}

\begin{equation}
    \Xi(\mathbf r,\omega) = \int \frac{d^2 \mathbf k}{(2\pi)^2} \frac{\epsilon_{\alpha \beta}}{2}
 \langle \mathbf r  | \gvz (\partial_{k_\alpha}\gvz  ^{-1}) (\partial_{k_\beta}\gvz)| \mathbf r \rangle.
\end{equation}

where $\gvz$ is the valley-spin Green's function defined as

\begin{equation}
	\gvz (\omega) = [\omega - H(\mathbf k) +i 0^+]^{-1} \mathcal{V}_z S_z
\end{equation}
with $\mathcal{V}_z$ the valley operator,\cite{PhysRevLett.120.086603,PhysRevLett.121.146801,PhysRevB.99.245118,2020arXiv200305163M,2020arXiv200502169L} $S_z$ the spin operator
and $H(\mathbf k)$ the Bloch Hamiltonian.

The valley-spin Chern number
$\mathcal{C}$
can be computed as $\mathcal{C} = \int_{-\infty}^0 d\omega \int d^2 
\mathbf r \Xi(\mathbf r,\omega) $, and as result
$ \Xi(\mathbf r,\omega)$ the real space energy resolved
valley-spin flux.\cite{PhysRevB.84.205137,2020arXiv200305163M,PhysRevLett.123.096802,2020arXiv200502169L}
The valley operator $\mathcal{V}_z$ is defined in the tight binding 
basis as\cite{PhysRevLett.120.086603,PhysRevLett.121.146801,PhysRevB.99.245118,2020arXiv200305163M,2020arXiv200502169L}

\begin{equation}
	\mathcal{V}_z = 
	\frac{i}{3\sqrt{3}} \sum_{\langle \langle ij\rangle \rangle,s} 
	\vartheta^z_{ij} \nu_{ij}
        c^\dagger_{i,s} c_{j,s}
	\label{eq:valley}
\end{equation}
where $\nu_{ij}=\pm 1$ for clock-wise/anticlock-wise hopping, 
$\langle \langle \rangle \rangle$ denotes sum over second neighbors
in a layer, and
$\vartheta^z_{ij}$ is the sublattice Pauli matrix. The previous operator takes
the eigenvalues $\pm 1$ at the microscopic $K$ and $K'$, and as a result
can be used to compute the valley flavor of an eigenstate. it
is worth to note that this operator
will be especially useful in the next section when computing valley symmetry
breaking in the real tight binding model.

Focusing on the case with antiferromagnetic configuration, we
observe that real space valley fluxes
at the Fermi energy $\omega=\epsilon_F$ emerge in the unit cell
(Fig. \ref{fig:fig3}b),
similarly to other van der Waals multilayers.\cite{PhysRevLett.123.096802,2020arXiv200502169L,2020arXiv200305163M}
The combination of triangular arrangement (Fig. \ref{fig:fig3}a)
and real space valley-spin flux (Fig. \ref{fig:fig3}b)
suggest that the low energy band structure can be captured with
an effective triangular lattice with spin-valley fluxes
$
	H = \sum_{\langle ij \rangle,s,s'} e^{i \sigma^z_{s,s'} \varphi_{ij}}
	d^\dagger_{i,s} d_{j,s'} 
	$, 
	where $d^\dagger_{i,s},d_{i,s}$ are the creation/annihilation operators
for the Wannier states, 
$\varphi_{ij}$ are the phases associated to the real space flux
of the supercell $\phi_S$ (Fig. \ref{fig:fig3}d). 
This low energy Wannier model would allow to study
interaction effects without having to perform a calculation
for the whole moire unit cell, and similar models have been
explored in a variety of twisted two dimensional materials.\cite{PhysRevLett.121.087001,PhysRevLett.123.096802,2020arXiv200313690S}
The form of the effective interactions in these low energy
models can be non-trivial, due to the potentially
extended nature of the Wannier states.\cite{PhysRevX.8.031087,PhysRevX.8.031088}
Here, in order to perform unbiased calculations,
we will explore interaction effects with the full
microscopic model including onsite, first and second neighbor
interactions.

\section{Interaction-induced symmetry breaking in exchanged-bias
flat bands}
\label{sec:int}
We now focus on the effect of interactions in the full real space model.
For that purpose, we consider a generic interacting Hamiltonian
with interactions up to second neighbors of the form

\begin{figure}[t!]
\centering
    \includegraphics[width=\columnwidth]{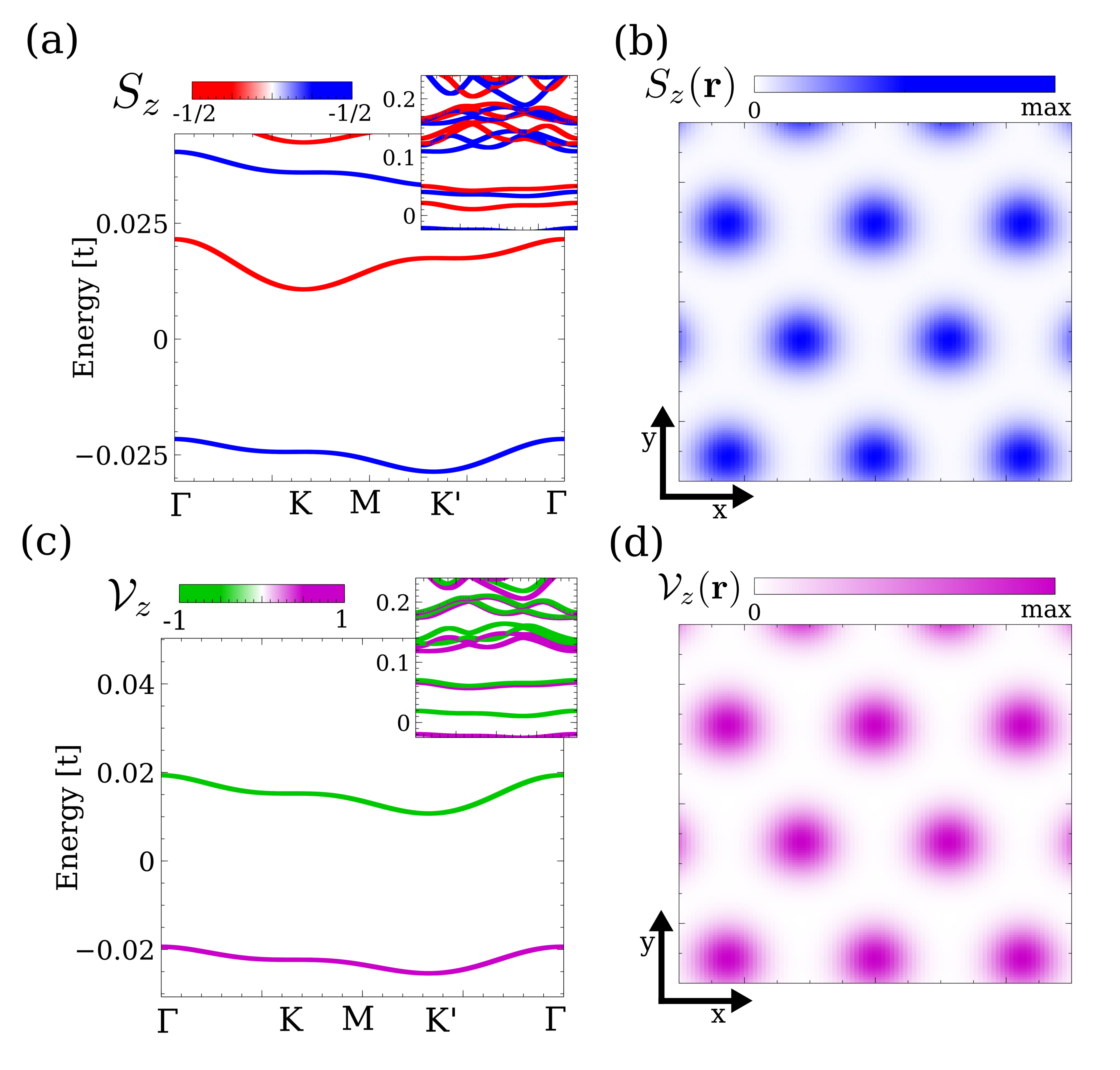}

\caption{Selfconsistent electronic structure with an
	antiferromagnetic alignment between the leads (a,b), and
	with a ferromagnetic one (c,d).
	In the case of an antiferromagnetic alignment (a,b),
	interactions give rise to a spin polarized ground state (a),
	with a net $S=1/2$ magnetic moment per unit cell (b).
	In the case of ferromagnetic alignment (c,d),
	interactions give rise to a valley
	polarized ground state (c), with a net valley
	moment per unit cell (d).
}
\label{fig:fig4}
\end{figure}

\begin{equation}
	H_I = 
	H_I^{(0)} + 
	H_I^{(1)} + 
	H_I^{(2)}
\end{equation}

where $H_I^{(0)}$ is the local Hubbard interaction with strength $U$

\begin{equation}
	H_I^{(0)} = U
	\sum_i c^\dagger_{i,\uparrow}c_{i,\uparrow}
	c^\dagger_{i,\downarrow}c_{i,\downarrow}
	\label{v1}
\end{equation}

$H_I^{(1)}$ is the first neighbor interaction with strength $V_1$

\begin{equation}
	H_I^{(1)} =
	V_1
	\sum_{\langle ij \rangle} \left [
	\left ( \sum_s c^\dagger_{i,s}c_{i,s}  \right )
	\left ( \sum_s c^\dagger_{j,s}c_{j,s}  \right )
	\right ]
	\label{v2}
\end{equation}

$H_I^{(2)}$ is the second neighbor interaction with strength $V_2$

\begin{equation}
	H_I^{(2)} =
	V_2
	\sum_{\langle \langle ij \rangle \rangle} \left [
	\left ( \sum_s c^\dagger_{i,s}c_{i,s}  \right )
	\left ( \sum_s c^\dagger_{j,s}c_{j,s}  \right )
	\right ]
	\label{v3}
\end{equation}

We note that by construction, the interaction Hamiltonian is
SU(2) symmetric. We decouple the previous Hamiltonian
with a conventional non-collinear mean-field approximation,
which we solve selfconsistently. In the following we take
$U=2t$, $V_1 = 1.5t$ and $V_2 = 0.5t$, and we will
focus on the the of a single electron per moire
unit cell, that turns the low energy flat bands
half filled. We note that the same results are expected for
a single hole per unit cell and, as it will be shown later,
this would be physically achievable regime for the
WSe$_2$/CrBr$_3$ twisted multilayer.

We now consider the
two different scenarios for the magnetic encapsulation, 
the heterostructure with with antiferromagnetic
configuration (Fig. \ref{fig:fig4}ab), 
and with ferromagnetic configuration (Fig. \ref{fig:fig4}cd).
As elaborated above, depending on the
configuration of the magnetic electrodes, the spin or valley degeneracy
is lifted. In the low energy manifold, which is now just two fold degenerate,
interactions can give rise to an additional symmetry breaking.

We focus first on the case with antiferromagnetic configuration,
whose low energy manifold has two bands, one per spin channel
(Fig. \ref{fig:fig2}f).
Due to this spin degeneracy, correlation effects could be able
to lift the degeneracy and open up a gap, as we explicitly show below.
We now take the interaction terms introduced in 
Eqs. \ref{v1}\ref{v2}\ref{v3}, and solve the selfconsistent mean
field equations. We obtain that, 
upon introducing interactions,
symmetry broken state emerges, hosting a net
spin magnetic moment per supercell
and lifting the degeneracy of the low energy band structure.\cite{PhysRevB.98.161406}
The mean field
band structure shown in Fig. \ref{fig:fig4}a shows that an spontaneous
exchange splitting appears in the band structure,
lifting the degeneracy of the low energy manifold
introduced in Fig. \ref{fig:fig2}f.
The projection in real space of the
spin operator $S_z (\mathbf r)$ 
shows that, associated with the lifting of degeneracy,
a non-zero magnetic moment emerges, localized in the
regions with AB stacking as shown in Fig. \ref{fig:fig4}b.

We now move on to the case in which the magnetic leads
have a ferromagnetic alignment. In this situation, the low energy
manifold host two bands, one per valley (Fig. \ref{fig:fig2}e).
Given that these two energy band belong to the same spin channel,
a magnetic instability cannot lift their degeneracy, in stark
contrast with the antiferromagnetic encapsulation.
Upon introducing electronic interactions of Eqs. \ref{v1}\ref{v2}\ref{v3}
and solving the selfconsistent problem, we obtain
a mean field Hamiltonian whose
band-structure has an spontaneous
interaction induced valley splitting
(Fig. \ref{fig:fig4}c). 
We can compute the expectation value of the valley operator
in the unit cell $\mathcal{V}_z (\mathbf r)$, and we
observe that a non-zero valley polarization emerges
in the regions with AB stacking as shown in 
Fig. \ref{fig:fig4}d. The real space expectation value
of the valley operator is computed by taking the valley operator
of Eq. \ref{eq:valley} and adding a localized real space envelope
to each site.

The previous calculations demonstrate, using a microscopic interacting model,
that the symmetry breaking induced by interactions is controlled
by the magnetic alignment of the magnetic substrate.
Importantly, the genuine combination of spin-orbit and exchange effects
lead to the realization of pure-valley or pure-spin electronic
instabilities in the flat bands of twisted dichalcogenides.
Finally, it is worth to mention that the previous discussion has focused
in a low energy atomistic effective model. In the next section we show
using first principles methods, that a specific van der materials
formed by WSe$_2$/CrBr$_3$ would be described with the effective model
considered.

\begin{figure}[t!]
\centering
    \includegraphics[width=\columnwidth]{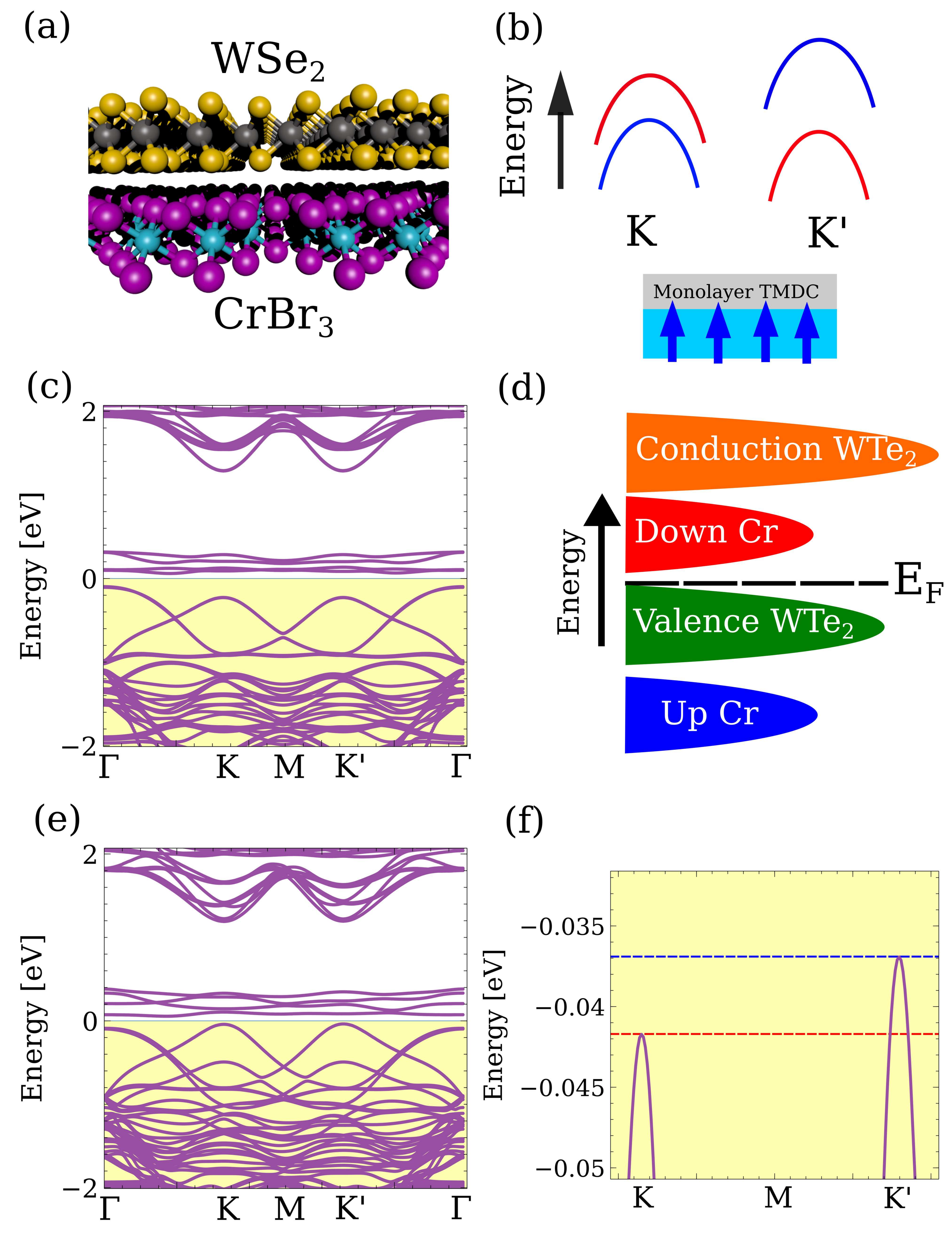}

\caption{
	(a) Sketch of the structure computed from first principles,
	consisting on WSe$_2$ on top of CrBr$_3$.
	Panel (b) shows a sketch of the effect
	of the exchange proximity effect,
	creating a splitting between the states at K an K'.
	Panel (c) shows the first principles band-structure
	of the heterostructure in (a), in the absence of spin-orbit
	coupling. Panel (d) shows a sketch of the orbitally
	resolved density of states as obtained in (c).
	Panel (e) shows the first principles
	band structure will full non-collinear
	spin-orbit coupling, and panel (f) shows a zoom
	on the top of the valence band.
}
\label{fig:fig5}
\end{figure}

\section{First principles calculations of the WSe$_2$/CrBr$_3$
exchange bias}
\label{sec:ex}

We finally show via
first principles calculations
that proximity effect between WSe$_2$ and CrBr$_3$ leads to
an exchange splitting in the valence band of WSe$_2$.
The emergence of such splitting
can be understood from a second order process in which an electron
jumps from the semiconductor to the ferromagnet and back.\cite{Klein2018,PhysRevLett.121.067701,Zhou2019,Mashhadi2019,PhysRevB.100.085128}
Since the emergence of flat bands in twisted dichalcogenides
have already been demonstrated by first principles, here
with will focus on the combined effect of exchange proximity
effect and spin-orbit coupling. For this sake, we consider a
minimal multilayer consisting on WSe$_2$ on top of the ferromagnetic
insulator CrBr$_3$, as shown in Fig. \ref{fig:fig5}a. 
In this situation, the exchange proximity effect created by CrBr$_3$
on WSe$_2$ will give rise to valley splitting in the
low energy band-structure, as sketched in Fig. \ref{fig:fig5}b.

We performed our first principles calculations with the all-electron
LAPW formalism as implemented in Elk.\cite{elk} Correlations in the Cr d-manifold
where included with the DFT+U formalism in the Yukawa form,\cite{PhysRevB.80.035121} taking
$U=4$ eV. 
We first focus on the case without spin-orbit
coupling, the resulting band structure is shown
in Fig. \ref{fig:fig5}c.
The first important feature is that,
in the absence of spin-orbit coupling, the top of the valence band is
located at $\Gamma$ instead of $K$, and as will be shown later this
dramatically changes when spin-orbit coupling is included. Second,
it is also worth to note that the conduction band states are formed
by unoccupied Cr orbitals, instead of the WSe$_2$ 
states. This orbital arrangement
as a function of energy is shown
in Fig. \ref{fig:fig5}d. This indicates
that for the current van der Waals heterostructure, correlated
states in moire flat bands should be searched in the hole-doped regime.
Although it is well known that DFT underestimates the Mott
gap in correlated magnets, our calculations performed with $U=4$ eV
for the Cr d-orbitals suggest that they might indeed be located inside
the gap in real multilayer.
Nevertheless, it must be noted that using a different 
ferromagnetic insulator such as EuS$_2$ may allow
to explore the effect of exchange field in conduction electron flat bands.

We now move to consider the heterostructure computed with full
non-collinear spin-orbit coupling, whose band structure is shown
in Fig. \ref{fig:fig5}e. As anticipated before, in the presence
of spin-orbit coupling the top of the valence band is located at
$K$, in agreement with the low energy model used before. This feature
highlights that first principles calculations
of flat bands in twisted WSe$_2$ can be strongly impacted by the
presence of spin orbit coupling,
as it dramatically impact the nature of the states at the top of the
valence band. 
In particular we find that,
the top of the valence band is 60 meV below the top of the
valence band at the $K$/$K'$ point. 
Second, it is observed that the combination
of magnetism and spin-orbit coupling
breaks the degeneracy between $k \rightarrow -k$, giving rise
to different dispersion around the $K$ and $K'$ points (Fig. \ref{fig:fig5}e).
In the present case, we are particularly interested on its impact at the
top of the
valence band.
Zooming the band structure on the top of the valence band as shown in
Fig. \ref{fig:fig5}f,
we clearly observe a splitting between
the K and K' point, which is exactly associated to
the combination of exchange
proximity effect between CrBr$_3$ and WSe$_2$
and spin-orbit coupling.\cite{PhysRevB.100.085128,Catarina2020,PhysRevLett.124.197401,2020arXiv200404073L}
The exchange proximity effect is found
on the order of 4 meV, similar to other van der 
Waals multilayers.\cite{PhysRevB.100.085128}
The previous first principles results demonstrate both the validity of the
low energy model used previous for the top of the valence band, and
the possibility of breaking $K/K'$ via exchange proximity effect,
the feature that we used too selectively lift the degeneracy
of the flat bands.

We finally note that the previous first principles calculations
do not account from additional effects appearing in the twisted system,
such as atomic relaxations in the twisted WSe$_2$, rippling between 
the WSe$_2$ and CrBr$_3$, or atomic relaxations in CrBr$_3$ stemming from
the moire pattern between CrBr$_3$ and WSe$_2$. To capture those
effects, first principles calculations of a twisted 
CrBr$_3$/WSe$_2$/WSe$_2$/CrBr$_3$ multilayer should be performed,
including non-collinear spin-orbit coupling and DFT+U. 
Nevertheless, given the large unit cells involved, this would
be represent a challenging system from the computation point of view.

\section{Conclusions}
\label{sec:con}
We have shown that a van der Waals heterostructure
consisting of twisted dichalcogenide bilayer (WSe$_2$) 
encapsulated
between two-dimensional ferromagnets (CrBr$_3$) shows flat bands and
correlated states controlled by the magnetic encapsulation. 
By using an effective atomistic model that incorporates
spin-orbit and exchange proximity effects,
we showed that such moire system shows flat bands
whose internal orbital structure strongly depends
on the magnetic encapsulation.
Once we include electronic interactions,
we demonstrated that depending on the magnetic 
arrangement, a spontaneous spin-symmetry breaking or
valley-symmetry breaking emerges
due to correlation effects.
Finally, using first-principles calculations, we showed that
the proximity between CrBr$_3$
and WSe$_2$ creates a spin splitting in the WSe$_2$,
in agreement with the low energy model used.
Our results put forward hybrid magnetic/dichalcogenide twisted
multilayers as a versatile platform
where spin-orbit and exchange effects drive controllable correlated
states.

\section*{Acknowledgments}
D.S.
thanks financial support from EU through the MSCA
project Nr. 796795 SOT-2DvdW.
J.L.L. acknowledges the computational 
resources provided by the Aalto Science-IT project.
We thank Z. Sun, T. Wolf, G. Blatter, O. Zilberberg,
F. Guinea, B. Amorim, M. Sigrist, M. Roesner and P. Liljeroth 
for fruitful discussions.
Part of this work was carried out on the Dutch national e-infrastructure with
the support of SURF Cooperative

\bibliographystyle{apsrev4-1}
\bibliography{biblio}{}

\end{document}